\documentclass[prl,reprint,superscriptaddress,amsmath,amssymb,floatfix,twocolumn]{revtex4-2}

\usepackage{graphicx}
\usepackage{epstopdf}
\usepackage{epsfig}
\usepackage{wrapfig}
\usepackage{enumerate}

\DeclareGraphicsExtensions{.ps}

\begin{document}
\title{Electronic correlations and Fermi liquid behavior of intermediate-band states in  titanium-doped silicon}

\author{A. \"Ostlin}
\affiliation{Theoretical Physics III, Center for Electronic
Correlations and Magnetism, Institute of Physics, University of
Augsburg, D-86135 Augsburg, Germany}
\affiliation{Augsburg Center for Innovative Technologies, University of Augsburg,
D-86135 Augsburg, Germany}
\author{L. Chioncel}
\affiliation{Augsburg Center for Innovative Technologies, University of Augsburg,
D-86135 Augsburg, Germany}
\affiliation{Theoretical Physics III, Center for Electronic
Correlations and Magnetism, Institute of Physics, University of
Augsburg, D-86135 Augsburg, Germany}
 
\begin{abstract} 
We study the nature of the electronic states in the intermediate band formed by interstitial titanium in silicon. Our single-site description combines effects of electronic correlations, captured by dynamical mean-field theory, and disorder, modeled using the coherent potential approximation and the typical medium mean-field theory. For all studied concentrations an extended metallic state with a strongly depleted density of states at the Fermi level is obtained. The self-energy is characteristic to Fermi-liquids and for certain temperatures reveals the existence of coherent quasi-particles.
\end{abstract}

\maketitle 
The metal-to-insulator transition (MIT) is a rich field of research within condensed-matter physics, where many different underlying mechanisms (disorder and/or interaction) has been proposed~\cite{im.fu.98,vlad.12}. 
Anderson~\cite{anderson.58} introduced a mechanism which now bears his name, while studying doping of phosphor in silicon, where the transition is due to the randomness of the impurities. 
At about the same time, Mott~\cite{mott.68} proposed that screening effects would become more important as the number of impurities increases, leading to a MIT. A third possibility is the Mott-Hubbard transition~\cite{im.fu.98}, in which an on-site intra-atomic electron-electron Coulomb interaction competes with inter-atomic hopping, leading potentially to an insulating state. 
In principle, all three of these mechanisms are present in doped semiconductors, and to disentangle one from the other is a non-trivial task.

From the family of doped semiconductors, one promising candidate for the next generation of solar cells are the intermediate-band (IB) photovoltaics~\cite{lu.ma.97,lu.ma.10,lu.ma.12}. In this class of materials, deep-level impurities are used to create a narrow
band situated inside the gap of a semiconductor host. The IB allows two-photon absorption processes, which are beneficial for the conversion efficiency of solar-cell devices.
However, the deep-level impurity may also give rise to a localized charge around itself, which in turn distorts the lattice. 
This can lead to non-radiative recombination processes, as discussed by Shockley, Reed~\cite{sh.re.52}, and Hall~\cite{hall.52}, where the efficiency is reduced due to phonons.
It has been argued that the non-radiative recombination can be counteracted by increasing the doping of impurities to the point where the states in the IB becomes delocalized, making the charge
spread out~\cite{la.he.75,lu.ma.06}. Therefore, it is an important question to answer if the IB states are localized or delocalized, or in other words, if the system is insulating or metallic.

An often studied IB photovoltaic is silicon doped with titanium~\cite{an.ma.09,ol.lo.16,wa.be.21}. During the last two decades, there has been a substantial research interest in this system, with conflicting results and interpretations. 
It seems that both the crystal growth techniques used to grow the silicon host, and the implantation method for the Ti, play important roles.
Both experiment and theory indicates that Ti impurities predominantly occupy interstitial positions
in the diamond structure of Si~\cite{lu.ma.06,sa.ag.09,li.wa.18,li.ak.21}. Since the solubility limit of Ti in Si is relatively low~\cite{ho.ma.88,ma.ba.91}, out-of-equilibrium implantation techniques have to be
used to achieve so-called hyperdoping. The Mott limit, where the states in the IB would turn from localized to delocalized, is estimated to be $N_c = 6 \times 10^{19}$ cm$^{-3}$~\cite{lu.ma.06}, 
therefore hyperdoping is necessary to reach the MIT.
High Ti concentrations have been reported, some showing as high as $10^{19}$ and $10^{21}$ cm$^{-3}$~\cite{an.ma.09,ol.pa.12}. However, even though good conductivity and sub-bandgap photoresponse has been demonstrated, the specific detectivity was found to be rather low~\cite{wa.be.21}.
Liu \emph{et al.}~\cite{li.wa.18} used two different Ti-implantation techniques, where one sample was completely insulating for all concentrations, while the other had a decreasing sheet resistance with increasing Ti concentration. In the latter case, the decrease in conductivity was attributed to percolation in Ti-rich cellular walls, and not to the MIT. This brings into question the degree of spatial homogeneity of Ti one can expect after hyperdoping. Recent studies, using ion implantation and pulsed-laser melting, came to the conclusion that the possible concentrations are well below the Mott limit~\cite{li.ak.21,ak.ma.21}.

On the theoretical side, in an attempt to formulate a mean-field theory of Anderson localization involving one-particle quantities, it was realized that the typical density of states (TDOS) vanishes as the disorder strength increases, and therefore can serve as a possible order parameter for localization~\cite{do.ko.97,do.pa.03,dobr.10}. 
Such an effect can not be captured by using the arithmetic average of the self-consistent coherent potential approximation (CPA)~\cite{tayl.67,sove.67}.
Within the proposed typical medium theory (TMT)~\cite{do.pa.03}, geometric averaging over disorder was used, instead of the arithmetic average.
A combined density functional theory (DFT)~\cite{jones.15} calculation followed by the cluster extension of the TMT, the so-called typical medium dynamical cluster approximation
TM-DCA~\cite{ek.te.14,zh.te.15}
recently reported that in the concentration range of $1.0 \times 10^{20}$ to $5.0 \times 10^{19}$ cm$^{-3}$ Ti impurities in Si~\cite{zh.ne.18} the IB electronic states within the gap undergo a localized to extended transition. 

As the incomplete $3d$ shell of Ti is subject to important electronic correlation effects, we investigate here the competition of electron-electron interaction and localization of IB electronic states in Ti-doped Si.  
On the methodological side we report the combined implementation of TMT and dynamical mean-field theory (DMFT)~\cite{me.vo.89,ko.vo.04}, to describe electronic correlation, into the DFT framework. 
This is a natural extension of a previous single-site TMT formulation using the exact muffin-tin orbitals (EMTO) basis set~\cite{os.zh.20}, which however departs from the original single-site TMT formulation as will be explained below.
We compare spectral functions of DFT(CPA/TMT)~\cite{os.zh.20}, with and without electronic correlation~\cite{os.vi.18}, with our present TMT+DMFT implementation.  
According to the present results the system remains metallic for the experimentally relevant Hubbard $U$ and $J$ parameters with a significantly reduced density of states at the Fermi level and a non-zero but reduced TMT order parameter.

\begin{figure}[h]
    \centering
    \includegraphics[width=\linewidth]{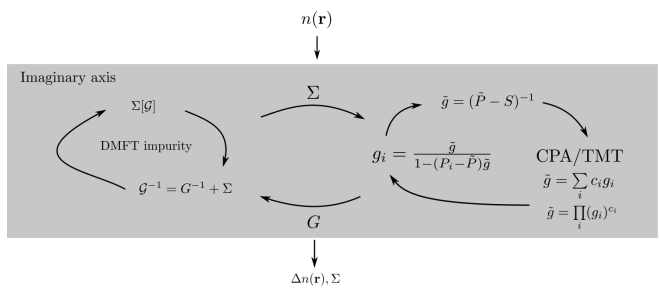}
    \caption{CPA/TMT+DMFT flow diagram. The path operator $\tilde{g}$ for the effective medium is arithmetically averaged in single-site CPA, while it is geometrically averaged in single-site TMT. The potential parameters $P$ and structure constants $S$ 
    parametrize the Green's function~\cite{os.vi.18}. The full charge self-consistency is achieved in the DFT external loop: the charge density $n({\bf r})$ is passed into the combined disorder and many-body solver, while the self-energy $\Sigma$ is stored for the next many-body iteration, and the correction $\Delta n({\bf r})$ to the real space charge density is returned into the outside DFT loop.}
    \label{fig:flow}
\end{figure}

In order to clarify the implementation let us state the key idea of the TMT namely that the typical values of the random quantities should be associated with physical observables~\cite{anderson.58}. The typical value of a variable $X$ is represented by its geometric mean $\langle X \rangle_{\mathrm{g}} \equiv \mathrm{exp} \ [ \langle \mathrm{ln} X(\varepsilon)\rangle_{\mathrm{a}}]$,
where $X(\varepsilon)$ is a function of the random on-site energy $\varepsilon$ and by $\langle ...\rangle_{\mathrm{a}}$ we denote the arithmetic average. The typical density of states reads: $\rho_g(E)$=$\mathrm{exp} \ [ \langle \ln \rho(E,\varepsilon)\rangle_{\mathrm{a}}]$. The typical value of the local Green's function is obtained as a Hilbert transform:
$G_g(z)=\int_{-\infty}^\infty \rho_g(E)/(z-E) \mathrm{d}E$, which in many situations lead to numerical difficulties.
In our TMT formulation we geometrically average the full scattering path operator ($\tilde{g}$, see Fig.~\ref{fig:flow}) in such a way we avoid the numerical instabilities associated with the Hilbert transform.
We follow the implementation of Ref.~\cite{os.vi.18}, which parametrize the Green's function (in terms of potential parameters $P$ and structure constants $S$) and thus does not require any analytical continuation within the DFT+DMFT self-consistency loops~\cite{os.vi.17}.
The schematic flow diagram is shown in Fig.~\ref{fig:flow} and follows that presented in Ref.~\cite{os.vi.18}. 

In order to solve the DMFT impurity problem, the perturbative spin-polarized $T$-matrix fluctuation-exchange (SPTFLEX) solver~\cite{bi.sc.89,li.ka.98,ka.li.99,po.ka.05,gr.ma.12} is used. 
In this solver, the electron-electron interaction term can be considered in a full spin and orbital rotationally invariant form, viz. $\frac{1}{2}\sum_{{i \{m, \sigma \} }} U_{mm'm''m'''}
c^{\dag}_{im\sigma}c^{\dag}_{im'\sigma'}c_{im'''\sigma'}c_{im''\sigma} $.
Here, $c_{im\sigma}/c^\dagger_{im\sigma}$ annihilates/creates an electron with 
spin $\sigma$ on the orbital $m$ at the lattice site $i$.
The Coulomb matrix elements $U_{mm'm''m'''}$ are expressed in the usual
way in terms of Slater integrals~\cite{im.fu.98}.
The simplifications of the computational procedure in reformulating the 
FLEX~\cite{bi.sc.89} as a DMFT impurity solver~\cite{li.ka.98,ka.li.99,po.ka.05,ka.ir.08} consists in neglecting dynamical interactions 
in the particle-particle channel, considering only static (of $T-$matrix type)
renormalization of the effective interactions. 
As some interaction effects are already
included in the DFT exchange-correlation functional, the so-called
``double-counted'' terms must be subtracted. To achieve this, we employ the fully-localized limit double-counting scheme~\cite{pe.ma.03}.
The Matsubara frequencies $i\omega_n=(2n+1)i\pi T$, with $n=0,1,2, \cdots$, were truncated after 1024 frequencies, and different values for the temperature $T$ were considered.
The densities of state were computed along a horizontal contour shifted
away from the real-energy axis by about $10$~meV. At the end of the self-consistent DFT calculations, the self-energy $\Sigma(i\omega_n)$ was analytically continued by a Pad\'e approximant~\cite{vi.se.77,os.ch.12} to the horizontal contour.

Silicon crystallizes in the diamond structure (space group 227, Fd$\overline{3}$m), with a lattice constant $a=5.4309$~{\AA}. The Si atoms occupy the $(0, 0, 0)$ and $a(\frac{1}{4}, \frac{1}{4}, \frac{1}{4})$ positions, while the $a(\frac{1}{2}, \frac{1}{2}, \frac{1}{2})$ and $a(\frac{3}{4}, \frac{3}{4}, \frac{3}{4})$ are unoccupied. We denote the unoccupied positions as Em (for ``empty") and consider the Ti-interstitial impurities as randomly placed at the $a(\frac{1}{2}, \frac{1}{2}, \frac{1}{2})$ lattice-sites. We view the Ti and the Em as forming a ``binary alloy", occupying the $a(\frac{1}{2}, \frac{1}{2}, \frac{1}{2})$-site with the formula Ti$_c$Em$_{1-c}$, where $0 \leq c \leq 1$ denotes the concentration of Ti. Throughout the paper, unless stated otherwise,
we use $c=0.001$, since this is a representative concentration close to the predicted Mott limit~\cite{lu.ma.06}.
We use the experimental lattice constant throughout all calculations, since
for the investigated concentrations of Ti impurities the change in unit cell volume upon doping is negligible~\cite{sa.ag.09,ma.sa.15}.
The single-site method that we use here can not, in general, capture the effect of local lattice relaxations. However, in previous studies the position of the IB was found to be fairly unaffected by the local lattice relaxation due to the Ti-impurities~\cite{sa.ag.09,zh.ne.18}.
The Green's function is computed for 16 energy points along a semicircular contour with a diameter of 1 Ry,
with the basis consisting of $s$-, $p$-, and $d$-orbitals.
The DFT convergence is achieved already at about $5 \cdot 10^3$ {\bf k}-points however DOS plots were produced using an integration over the irreducible Brillouin zone a mesh consisting of more than $10^4$ {\bf k}-points. 
The generalized gradient approximation (GGA) with the Perdew-Burke-Ernzerhof (PBE) parametrization~\cite{pe.bu.96} of the exchange-correlation functional was used throughout this paper. The GGA was used by S\'anchez \emph{et al.}~\cite{sa.ag.09} in studying Ti impurities in Si using supercells.

\begin{figure}[h]
    \centering
    \includegraphics[width=\linewidth]{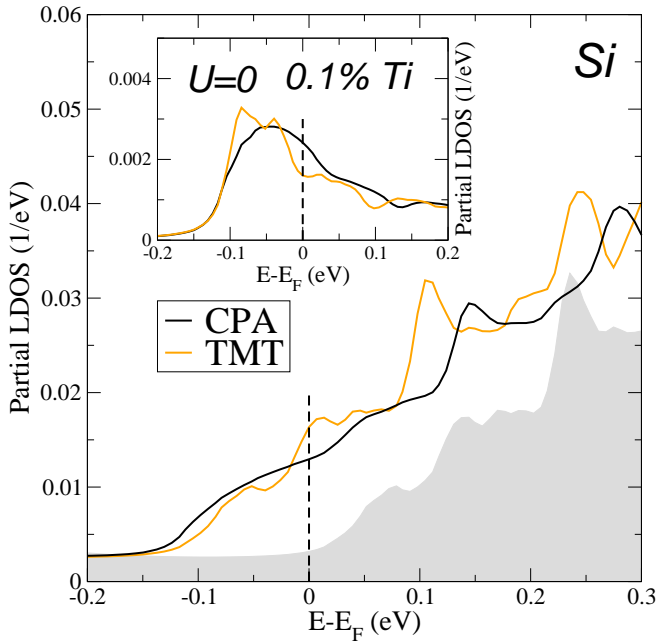}
    \caption{Concentration weighted Si partial DOS using CPA (black line) and TMT (orange line). Inset: Concentration-weighted partial DOS for interstitial Ti, at a concentration of $0.1\%$. The partial density of states of pure Si is shown as a shaded area.}
    \label{fig:u0dos}
\end{figure}

In Fig.~\ref{fig:u0dos} we show the partial LDOS for Si and Ti (inset) in the non-interacting case using the CPA and TMT configurational averages.
The IB states are formed by the hybridization of Ti-$3d$ states and the Si-$p$ states.
By orbital decomposition we confirm that at the Fermi level $E_F$, Ti-$t_{2g}$ orbitals are predominant. 
The impurity band in the CPA overlaps with the bottom conduction band, which is a consequence of the  poor performance of the traditional exchange correlation functional, which leads to an underestimated gap.
In general the arithmetic average is larger than the geometric one~\cite{bullen.03}, however for such a small concentration the average LDOS for the CPA and TMT are quite similar.
This might be attributed to the fact that Ti impurities are subject to similar hybridization with the tails of the Si-atoms on the vacated interstitial site.
We investigate further the behaviour of the TDOS, the so-called order parameter which monitors the presence of precursors for Anderson localization~\cite{sc.sc.10,do.pa.03,dobr.10}. In the absence of electronic correlations this quantity is computed according to: $\rho_g(E)=\rho^{c}_{\mathrm{Ti}}(E)\rho^{1-c}_{\mathrm{Em}}(E)$, and represents the geometrical average of the Ti-LDOS and the LDOS produced by all other atoms (mainly Si) at the position of the empty-site. Its magnitude remains essentially unmodified $\rho_g(E_F) \approx 0.008$~(eV)$^{-1}$ in the entire range of studied concentrations, and thus its non-zero value signals the lack of a disorder-driven MIT.

The effects of the mean-field decoupling of the Hubbard interaction has been addressed using the DFT+$U$ method~\cite{ma.le.14}.
A Hubbard correction to the $3d$ orbitals in the range of $U \approx 3.0 - 5.0$~eV give a good agreement for the magnitude of the migration barriers and the position of donor levels for certain transition metal impurities~\cite{ma.le.14,ma.sa.15} including Ti.
Given the spread in the magnitude of the $U$ parameter we have performed (CPA/TMT)+DMFT computations for several $U$ and $J$.

\begin{figure}[h]
    \centering
    \includegraphics[width=\linewidth]{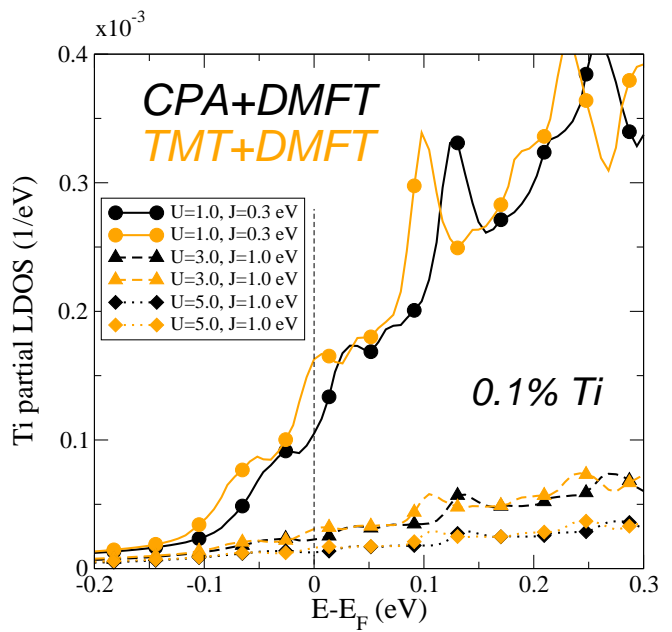}
    \caption{Ti partial density of states, for a doping level of $c=0.1\%$. CPA+DMFT (black lines) and TMT+DMFT (orange lines) for different values of the Coulomb $U$ and $J$ parameters for $T=400$~K.}
    \label{fig:us3}
\end{figure}

In Fig.~\ref{fig:us3} we show the Ti partial LDOS for different averaging schemes, and for different $U$ and $J$ parameters, at $T=400$~K. 
Electronic correlation effects on the Ti-$3d$ states lead to a spectral weight transfer towards the bottom of the conduction band (towards higher energies). Most significantly, depending on the strength of Hubbard parameters a major reduction of about two orders of magnitude of the Ti-LDOS at the Fermi level is obtained in the range of $\rho_{\mathrm{Ti}}(E_F)\approx 10^{-3}$ to $10^{-5}$ states/eV. 
Nevertheless the total DOS, $\rho(E_F)$, is not zero but with an absolute magnitude of about $10^{-2}$ states/eV. 
The CPA+DMFT methodology allows also to study the temperature dependence of the partial LDOS at the impurity site. For $U=5.0$~eV, $J=1.0$~eV  we found that in the vicinity of the Fermi level Ti-LDOS decreases with temperature,  however its magnitude is significantly smaller than the corresponding Em-LDOS (not shown).
Overall no significant temperature dependence of the total DOS at $E_F$ is obtained. 

\begin{figure}[h]
    \centering
        \includegraphics[width=\linewidth]{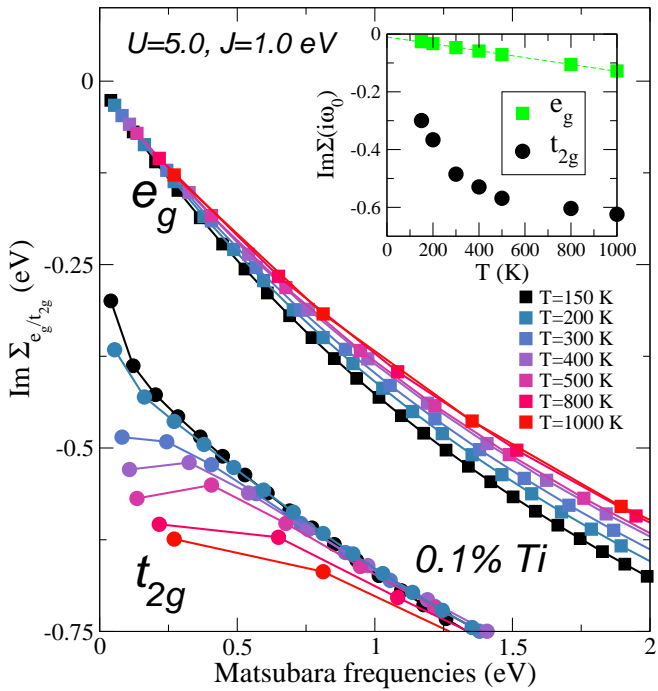}
    \caption{Temperature dependence of the orbital resolved Im $\Sigma_{e_g/t_{2g}}(i\omega_n)$ within the CPA+DMFT and at $U=5.0$~eV, J=1.0~eV.
    Squares/circles denote the $e_g/t_{2g}$-orbitals character. Lines are guides to the eye. Inset: Im $\Sigma_{e_g/t_{2g}}(i\omega_0)$ plotted versus temperature. The dashed line is a linear fit to the data.}
    \label{fig:mats_temp}
\end{figure}

In metallic alloys the self-energy carries information about the relaxation processes.
In particular the imaginary part of the DMFT self-energy encodes the Coulomb scattering effects. In disordered systems in the absence of interactions the self-energy takes into account the fluctuations in the on-site energies beyond the effective medium fields of CPA and TMT, and carries the information about the scattering on impurities.
In the case of local Fermi liquids it was recently shown that even in the presence of a constant scattering rate (which may mimic impurity scattering) into the DMFT self-energy the system still displays a Fermi-liquid behavior~\cite{be.mr.13}.

In Fig.~\ref{fig:mats_temp} we show the Ti $3d$-orbital-resolved self-energies as functions of  the Matsubara frequencies from the CPA+DMFT computation. Similar results were obtained from TMT+DMFT.
The Fermi liquid state is characterized by a linear dependence of the imaginary part of the self-energy on imaginary frequencies $i\omega_n$:   $\mathrm{Im} \Sigma(i\omega_n)=-\Gamma-(Z^{-1}-1)\omega_n$, where $Z$ is the quasiparticle spectral weight and $\Gamma$ the quasiparticle damping (inverse quasiparticle life time).
The $\mathrm{Im} \Sigma_{e_g}(i\omega_n)$ approaches zero linearly for all temperatures in the region of low Matsubara frequencies, thus the $e_g$-electrons are long-lived quasiparticles, i.e. their scattering rate vanishes as the Fermi surface is approached. 
A downturn in the $\mathrm{Im} \Sigma_{t_{2g}}(i\omega_n)$ is seen in the temperature range from 1000~K down to 300~K.
However, for temperatures between $200\,\mathrm{K} \le T \le 300$~K, the $\mathrm{Im}  \Sigma_{t_{2g}}(i\omega_n)$ is changing its curvature, so one expects that in this temperature range a cross over to the Fermi liquid regime occurs. 
The inset of Fig.~\ref{fig:mats_temp} shows that Im $\Sigma_{e_g/t_{2g}}(i\omega_0)$ versus temperature can be fitted to a line. In view of the ``first Matsubara frequency'' rule~\cite{ch.ma.12} such a linear dependence provides an estimate for the crossover temperature below which quasiparticles are coherent. 
While $e_g$-electrons are coherent quasiparticles in the entire temperature range the $t_{2g}$-electrons behave as coherent quasiparticles only below $\approx200$~K. These results provide a clear indication of the Fermi-liquid character of the electronic system.
The quasiparticle mass enhancement $m^{\star}/m$, with respect to the band mass $m$, can be approximately computed from the Matsubara frequencies in the zero-temperature limit as $m^{\star}/m=1-\frac{\mathrm{Im}\Sigma(i\omega_n)}{\omega_n}|_{\omega_n\rightarrow 0}$.
For the Hubbard parameter $U=5$~eV, we estimate a mass enhancement of $m^{\star}/m\approx8$ for the $t_{2g}$ orbitals at $150$~K, which confirms the presence of important correlation effects.

In summary, we modeled randomness and local Coulomb interaction effects in titanium-doped silicon. 
We compare results of CPA+DMFT~\cite{os.vi.18} with the present TMT+DMFT implementation and address the controversy whether the induced IB states, formed as  consequence of Ti-$d$($t_{2g}$) and Si-$p$ orbital hybridization, are localized or extended.  
We demonstrate that within the single-site version of TMT-EMTO, in the absence of electronic correlations, the order parameter that describes the precursors of Anderson localization is non-zero, so that no transition can be captured.
In the presence of interactions we report a significant reduction of the one-particle spectra at $E_F$, however the electronic states remain coherent quasiparticles at least up to $200$~K, despite a significantly large Hubbard $U$ parameter.
The drastic reduction of $\rho(E_F)$ is a predominant correlation effect being qualitatively and even quantitatively similar between CPA+DMFT and TMT+DMFT.
Thus, if a metal to insulator transition exists this should be triggered
by effects which are beyond the combined single-site disorder and correlation considered here. 
Note that recent experimental findings, points towards granular and percolation physics being of importance in this system, which requires modeling beyond local theories.

We have benefited from discussions with H. Terletska, Y. Zhang and L. Vitos. Funded by the Deutsche Forschungsgemeinschaft (DFG, German Research Foundation) --- Grant number 107745057 --- TRR 80.  

\bibliography{ref}

\begin{thebibliography}{51}%
\makeatletter
\providecommand \@ifxundefined [1]{%
 \@ifx{#1\undefined}
}%
\providecommand \@ifnum [1]{%
 \ifnum #1\expandafter \@firstoftwo
 \else \expandafter \@secondoftwo
 \fi
}%
\providecommand \@ifx [1]{%
 \ifx #1\expandafter \@firstoftwo
 \else \expandafter \@secondoftwo
 \fi
}%
\providecommand \natexlab [1]{#1}%
\providecommand \enquote  [1]{``#1''}%
\providecommand \bibnamefont  [1]{#1}%
\providecommand \bibfnamefont [1]{#1}%
\providecommand \citenamefont [1]{#1}%
\providecommand \href@noop [0]{\@secondoftwo}%
\providecommand \href [0]{\begingroup \@sanitize@url \@href}%
\providecommand \@href[1]{\@@startlink{#1}\@@href}%
\providecommand \@@href[1]{\endgroup#1\@@endlink}%
\providecommand \@sanitize@url [0]{\catcode `\\12\catcode `\$12\catcode
  `\&12\catcode `\#12\catcode `\^12\catcode `\_12\catcode `\%12\relax}%
\providecommand \@@startlink[1]{}%
\providecommand \@@endlink[0]{}%
\providecommand \url  [0]{\begingroup\@sanitize@url \@url }%
\providecommand \@url [1]{\endgroup\@href {#1}{\urlprefix }}%
\providecommand \urlprefix  [0]{URL }%
\providecommand \Eprint [0]{\href }%
\providecommand \doibase [0]{https://doi.org/}%
\providecommand \selectlanguage [0]{\@gobble}%
\providecommand \bibinfo  [0]{\@secondoftwo}%
\providecommand \bibfield  [0]{\@secondoftwo}%
\providecommand \translation [1]{[#1]}%
\providecommand \BibitemOpen [0]{}%
\providecommand \bibitemStop [0]{}%
\providecommand \bibitemNoStop [0]{.\EOS\space}%
\providecommand \EOS [0]{\spacefactor3000\relax}%
\providecommand \BibitemShut  [1]{\csname bibitem#1\endcsname}%
\let\auto@bib@innerbib\@empty
\bibitem [{\citenamefont {Imada}\ \emph {et~al.}(1998)\citenamefont {Imada},
  \citenamefont {Fujimori},\ and\ \citenamefont {Tokura}}]{im.fu.98}%
  \BibitemOpen
  \bibfield  {author} {\bibinfo {author} {\bibfnamefont {M.}~\bibnamefont
  {Imada}}, \bibinfo {author} {\bibfnamefont {A.}~\bibnamefont {Fujimori}},\
  and\ \bibinfo {author} {\bibfnamefont {Y.}~\bibnamefont {Tokura}},\
  }\bibfield  {title} {\bibinfo {title} {Metal-insulator transitions},\ }\href
  {https://doi.org/10.1103/RevModPhys.70.1039} {\bibfield  {journal} {\bibinfo
  {journal} {Rev. Mod. Phys.}\ }\textbf {\bibinfo {volume} {70}},\ \bibinfo
  {pages} {1039} (\bibinfo {year} {1998})}\BibitemShut {NoStop}%
\bibitem [{\citenamefont {Dobrosavljevic}\ \emph {et~al.}(2012)\citenamefont
  {Dobrosavljevic}, \citenamefont {Trivedi},\ and\ \citenamefont
  {Valles~Jr.}}]{vlad.12}%
  \BibitemOpen
  \bibinfo {editor} {\bibfnamefont {V.}~\bibnamefont {Dobrosavljevic}},
  \bibinfo {editor} {\bibfnamefont {N.}~\bibnamefont {Trivedi}},\ and\ \bibinfo
  {editor} {\bibfnamefont {J.~M.}\ \bibnamefont {Valles~Jr.}},\ eds.,\ \href
  {https://doi.org/10.1093/acprof:oso/9780199592593.001.0001} {\emph {\bibinfo
  {title} {Conductor-Insulator Quantum Phase Transitions}}}\ (\bibinfo
  {publisher} {Oxford University Press},\ \bibinfo {address} {Oxford},\
  \bibinfo {year} {2012})\BibitemShut {NoStop}%
\bibitem [{\citenamefont {Anderson}(1958)}]{anderson.58}%
  \BibitemOpen
  \bibfield  {author} {\bibinfo {author} {\bibfnamefont {P.~W.}\ \bibnamefont
  {Anderson}},\ }\bibfield  {title} {\bibinfo {title} {Absence of diffusion in
  certain random lattices},\ }\href {https://doi.org/10.1103/PhysRev.109.1492}
  {\bibfield  {journal} {\bibinfo  {journal} {Phys. Rev.}\ }\textbf {\bibinfo
  {volume} {109}},\ \bibinfo {pages} {1492} (\bibinfo {year}
  {1958})}\BibitemShut {NoStop}%
\bibitem [{\citenamefont {Mott}(1968)}]{mott.68}%
  \BibitemOpen
  \bibfield  {author} {\bibinfo {author} {\bibfnamefont {N.~F.}\ \bibnamefont
  {Mott}},\ }\bibfield  {title} {\bibinfo {title} {Metal-insulator
  transition},\ }\href {https://doi.org/10.1103/RevModPhys.40.677} {\bibfield
  {journal} {\bibinfo  {journal} {Rev. Mod. Phys.}\ }\textbf {\bibinfo {volume}
  {40}},\ \bibinfo {pages} {677} (\bibinfo {year} {1968})}\BibitemShut
  {NoStop}%
\bibitem [{\citenamefont {Luque}\ and\ \citenamefont
  {Mart\'{\i}}(1997)}]{lu.ma.97}%
  \BibitemOpen
  \bibfield  {author} {\bibinfo {author} {\bibfnamefont {A.}~\bibnamefont
  {Luque}}\ and\ \bibinfo {author} {\bibfnamefont {A.}~\bibnamefont
  {Mart\'{\i}}},\ }\bibfield  {title} {\bibinfo {title} {Increasing the
  efficiency of ideal solar cells by photon induced transitions at intermediate
  levels},\ }\href {https://doi.org/10.1103/PhysRevLett.78.5014} {\bibfield
  {journal} {\bibinfo  {journal} {Phys. Rev. Lett.}\ }\textbf {\bibinfo
  {volume} {78}},\ \bibinfo {pages} {5014} (\bibinfo {year}
  {1997})}\BibitemShut {NoStop}%
\bibitem [{\citenamefont {Luque}\ and\ \citenamefont
  {Martí}(2010)}]{lu.ma.10}%
  \BibitemOpen
  \bibfield  {author} {\bibinfo {author} {\bibfnamefont {A.}~\bibnamefont
  {Luque}}\ and\ \bibinfo {author} {\bibfnamefont {A.}~\bibnamefont {Martí}},\
  }\bibfield  {title} {\bibinfo {title} {The intermediate band solar cell:
  Progress toward the realization of an attractive concept},\ }\href
  {https://doi.org/https://doi.org/10.1002/adma.200902388} {\bibfield
  {journal} {\bibinfo  {journal} {Advanced Materials}\ }\textbf {\bibinfo
  {volume} {22}},\ \bibinfo {pages} {160} (\bibinfo {year} {2010})}\BibitemShut
  {NoStop}%
\bibitem [{\citenamefont {Luque}\ \emph {et~al.}(2012)\citenamefont {Luque},
  \citenamefont {Mart{\'i}},\ and\ \citenamefont {Stanley}}]{lu.ma.12}%
  \BibitemOpen
  \bibfield  {author} {\bibinfo {author} {\bibfnamefont {A.}~\bibnamefont
  {Luque}}, \bibinfo {author} {\bibfnamefont {A.}~\bibnamefont {Mart{\'i}}},\
  and\ \bibinfo {author} {\bibfnamefont {C.}~\bibnamefont {Stanley}},\
  }\bibfield  {title} {\bibinfo {title} {Understanding intermediate-band solar
  cells},\ }\href {https://doi.org/10.1038/nphoton.2012.1} {\bibfield
  {journal} {\bibinfo  {journal} {Nature Photonics}\ }\textbf {\bibinfo
  {volume} {6}},\ \bibinfo {pages} {146} (\bibinfo {year} {2012})}\BibitemShut
  {NoStop}%
\bibitem [{\citenamefont {Shockley}\ and\ \citenamefont
  {Read}(1952)}]{sh.re.52}%
  \BibitemOpen
  \bibfield  {author} {\bibinfo {author} {\bibfnamefont {W.}~\bibnamefont
  {Shockley}}\ and\ \bibinfo {author} {\bibfnamefont {W.~T.}\ \bibnamefont
  {Read}},\ }\bibfield  {title} {\bibinfo {title} {Statistics of the
  recombinations of holes and electrons},\ }\href
  {https://doi.org/10.1103/PhysRev.87.835} {\bibfield  {journal} {\bibinfo
  {journal} {Phys. Rev.}\ }\textbf {\bibinfo {volume} {87}},\ \bibinfo {pages}
  {835} (\bibinfo {year} {1952})}\BibitemShut {NoStop}%
\bibitem [{\citenamefont {Hall}(1952)}]{hall.52}%
  \BibitemOpen
  \bibfield  {author} {\bibinfo {author} {\bibfnamefont {R.~N.}\ \bibnamefont
  {Hall}},\ }\bibfield  {title} {\bibinfo {title} {Electron-hole recombination
  in germanium},\ }\href {https://doi.org/10.1103/PhysRev.87.387} {\bibfield
  {journal} {\bibinfo  {journal} {Phys. Rev.}\ }\textbf {\bibinfo {volume}
  {87}},\ \bibinfo {pages} {387} (\bibinfo {year} {1952})}\BibitemShut
  {NoStop}%
\bibitem [{\citenamefont {Lang}\ and\ \citenamefont {Henry}(1975)}]{la.he.75}%
  \BibitemOpen
  \bibfield  {author} {\bibinfo {author} {\bibfnamefont {D.~V.}\ \bibnamefont
  {Lang}}\ and\ \bibinfo {author} {\bibfnamefont {C.~H.}\ \bibnamefont
  {Henry}},\ }\bibfield  {title} {\bibinfo {title} {{Nonradiative Recombination
  at Deep Levels in GaAs and GaP by Lattice-Relaxation Multiphonon Emission}},\
  }\href {https://doi.org/10.1103/PhysRevLett.35.1525} {\bibfield  {journal}
  {\bibinfo  {journal} {Phys. Rev. Lett.}\ }\textbf {\bibinfo {volume} {35}},\
  \bibinfo {pages} {1525} (\bibinfo {year} {1975})}\BibitemShut {NoStop}%
\bibitem [{\citenamefont {Luque}\ \emph {et~al.}(2006)\citenamefont {Luque},
  \citenamefont {Martí}, \citenamefont {Antolín},\ and\ \citenamefont
  {Tablero}}]{lu.ma.06}%
  \BibitemOpen
  \bibfield  {author} {\bibinfo {author} {\bibfnamefont {A.}~\bibnamefont
  {Luque}}, \bibinfo {author} {\bibfnamefont {A.}~\bibnamefont {Martí}},
  \bibinfo {author} {\bibfnamefont {E.}~\bibnamefont {Antolín}},\ and\
  \bibinfo {author} {\bibfnamefont {C.}~\bibnamefont {Tablero}},\ }\bibfield
  {title} {\bibinfo {title} {Intermediate bands versus levels in non-radiative
  recombination},\ }\href
  {https://doi.org/https://doi.org/10.1016/j.physb.2006.03.006} {\bibfield
  {journal} {\bibinfo  {journal} {Physica B: Condensed Matter}\ }\textbf
  {\bibinfo {volume} {382}},\ \bibinfo {pages} {320} (\bibinfo {year}
  {2006})}\BibitemShut {NoStop}%
\bibitem [{\citenamefont {Antolín}\ \emph {et~al.}(2009)\citenamefont
  {Antolín}, \citenamefont {Martí}, \citenamefont {Olea}, \citenamefont
  {Pastor}, \citenamefont {González-Díaz}, \citenamefont {Mártil},\ and\
  \citenamefont {Luque}}]{an.ma.09}%
  \BibitemOpen
  \bibfield  {author} {\bibinfo {author} {\bibfnamefont {E.}~\bibnamefont
  {Antolín}}, \bibinfo {author} {\bibfnamefont {A.}~\bibnamefont {Martí}},
  \bibinfo {author} {\bibfnamefont {J.}~\bibnamefont {Olea}}, \bibinfo {author}
  {\bibfnamefont {D.}~\bibnamefont {Pastor}}, \bibinfo {author} {\bibfnamefont
  {G.}~\bibnamefont {González-Díaz}}, \bibinfo {author} {\bibfnamefont
  {I.}~\bibnamefont {Mártil}},\ and\ \bibinfo {author} {\bibfnamefont
  {A.}~\bibnamefont {Luque}},\ }\bibfield  {title} {\bibinfo {title} {Lifetime
  recovery in ultrahighly titanium-doped silicon for the implementation of an
  intermediate band material},\ }\href {https://doi.org/10.1063/1.3077202}
  {\bibfield  {journal} {\bibinfo  {journal} {Applied Physics Letters}\
  }\textbf {\bibinfo {volume} {94}},\ \bibinfo {pages} {042115} (\bibinfo
  {year} {2009})},\ \Eprint
  {https://arxiv.org/abs/https://doi.org/10.1063/1.3077202}
  {https://doi.org/10.1063/1.3077202} \BibitemShut {NoStop}%
\bibitem [{\citenamefont {Olea}\ \emph {et~al.}(2016)\citenamefont {Olea},
  \citenamefont {L{\'{o}}pez}, \citenamefont {Antol{\'{\i}}n}, \citenamefont
  {Mart{\'{\i}}}, \citenamefont {Luque}, \citenamefont {Garc{\'{\i}}a-Hemme},
  \citenamefont {Pastor}, \citenamefont {Garc{\'{\i}}a-Hernansanz},
  \citenamefont {del Prado},\ and\ \citenamefont
  {Gonz{\'{a}}lez-D{\'{\i}}az}}]{ol.lo.16}%
  \BibitemOpen
  \bibfield  {author} {\bibinfo {author} {\bibfnamefont {J.}~\bibnamefont
  {Olea}}, \bibinfo {author} {\bibfnamefont {E.}~\bibnamefont {L{\'{o}}pez}},
  \bibinfo {author} {\bibfnamefont {E.}~\bibnamefont {Antol{\'{\i}}n}},
  \bibinfo {author} {\bibfnamefont {A.}~\bibnamefont {Mart{\'{\i}}}}, \bibinfo
  {author} {\bibfnamefont {A.}~\bibnamefont {Luque}}, \bibinfo {author}
  {\bibfnamefont {E.}~\bibnamefont {Garc{\'{\i}}a-Hemme}}, \bibinfo {author}
  {\bibfnamefont {D.}~\bibnamefont {Pastor}}, \bibinfo {author} {\bibfnamefont
  {R.}~\bibnamefont {Garc{\'{\i}}a-Hernansanz}}, \bibinfo {author}
  {\bibfnamefont {A.}~\bibnamefont {del Prado}},\ and\ \bibinfo {author}
  {\bibfnamefont {G.}~\bibnamefont {Gonz{\'{a}}lez-D{\'{\i}}az}},\ }\bibfield
  {title} {\bibinfo {title} {Room temperature photo-response of titanium
  supersaturated silicon at energies over the bandgap},\ }\href
  {https://doi.org/10.1088/0022-3727/49/5/055103} {\bibfield  {journal}
  {\bibinfo  {journal} {J. Phys. D: Appl. Phys.}\ }\textbf {\bibinfo {volume}
  {49}},\ \bibinfo {pages} {055103} (\bibinfo {year} {2016})}\BibitemShut
  {NoStop}%
\bibitem [{\citenamefont {Wang}\ and\ \citenamefont
  {Berencén}(2021)}]{wa.be.21}%
  \BibitemOpen
  \bibfield  {author} {\bibinfo {author} {\bibfnamefont {M.}~\bibnamefont
  {Wang}}\ and\ \bibinfo {author} {\bibfnamefont {Y.}~\bibnamefont
  {Berencén}},\ }\bibfield  {title} {\bibinfo {title} {Room-temperature
  infrared photoresponse from ion beam–hyperdoped silicon},\ }\href
  {https://doi.org/https://doi.org/10.1002/pssa.202000260} {\bibfield
  {journal} {\bibinfo  {journal} {physica status solidi (a)}\ }\textbf
  {\bibinfo {volume} {218}},\ \bibinfo {pages} {2000260} (\bibinfo {year}
  {2021})}\BibitemShut {NoStop}%
\bibitem [{\citenamefont {S\'anchez}\ \emph {et~al.}(2009)\citenamefont
  {S\'anchez}, \citenamefont {Aguilera}, \citenamefont {Palacios},\ and\
  \citenamefont {Wahn\'on}}]{sa.ag.09}%
  \BibitemOpen
  \bibfield  {author} {\bibinfo {author} {\bibfnamefont {K.}~\bibnamefont
  {S\'anchez}}, \bibinfo {author} {\bibfnamefont {I.}~\bibnamefont {Aguilera}},
  \bibinfo {author} {\bibfnamefont {P.}~\bibnamefont {Palacios}},\ and\
  \bibinfo {author} {\bibfnamefont {P.}~\bibnamefont {Wahn\'on}},\ }\bibfield
  {title} {\bibinfo {title} {Assessment through first-principles calculations
  of an intermediate-band photovoltaic material based on {T}i-implanted
  silicon: Interstitial versus substitutional origin},\ }\href
  {https://doi.org/10.1103/PhysRevB.79.165203} {\bibfield  {journal} {\bibinfo
  {journal} {Phys. Rev. B}\ }\textbf {\bibinfo {volume} {79}},\ \bibinfo
  {pages} {165203} (\bibinfo {year} {2009})}\BibitemShut {NoStop}%
\bibitem [{\citenamefont {Liu}\ \emph {et~al.}(2018)\citenamefont {Liu},
  \citenamefont {Wang}, \citenamefont {Berenc{\'e}n}, \citenamefont {Prucnal},
  \citenamefont {Engler}, \citenamefont {H{\"u}bner}, \citenamefont {Yuan},
  \citenamefont {Heller}, \citenamefont {B{\"o}ttger}, \citenamefont {Rebohle},
  \citenamefont {Skorupa}, \citenamefont {Helm},\ and\ \citenamefont
  {Zhou}}]{li.wa.18}%
  \BibitemOpen
  \bibfield  {author} {\bibinfo {author} {\bibfnamefont {F.}~\bibnamefont
  {Liu}}, \bibinfo {author} {\bibfnamefont {M.}~\bibnamefont {Wang}}, \bibinfo
  {author} {\bibfnamefont {Y.}~\bibnamefont {Berenc{\'e}n}}, \bibinfo {author}
  {\bibfnamefont {S.}~\bibnamefont {Prucnal}}, \bibinfo {author} {\bibfnamefont
  {M.}~\bibnamefont {Engler}}, \bibinfo {author} {\bibfnamefont
  {R.}~\bibnamefont {H{\"u}bner}}, \bibinfo {author} {\bibfnamefont
  {Y.}~\bibnamefont {Yuan}}, \bibinfo {author} {\bibfnamefont {R.}~\bibnamefont
  {Heller}}, \bibinfo {author} {\bibfnamefont {R.}~\bibnamefont {B{\"o}ttger}},
  \bibinfo {author} {\bibfnamefont {L.}~\bibnamefont {Rebohle}}, \bibinfo
  {author} {\bibfnamefont {W.}~\bibnamefont {Skorupa}}, \bibinfo {author}
  {\bibfnamefont {M.}~\bibnamefont {Helm}},\ and\ \bibinfo {author}
  {\bibfnamefont {S.}~\bibnamefont {Zhou}},\ }\bibfield  {title} {\bibinfo
  {title} {On the insulator-to-metal transition in titanium-implanted
  silicon},\ }\href {https://doi.org/10.1038/s41598-018-22503-6} {\bibfield
  {journal} {\bibinfo  {journal} {Scientific Reports}\ }\textbf {\bibinfo
  {volume} {8}},\ \bibinfo {pages} {4164} (\bibinfo {year} {2018})}\BibitemShut
  {NoStop}%
\bibitem [{\citenamefont {Lim}\ \emph {et~al.}(2021)\citenamefont {Lim},
  \citenamefont {Akey}, \citenamefont {Napolitani}, \citenamefont {Chow},
  \citenamefont {Warrender},\ and\ \citenamefont {Williams}}]{li.ak.21}%
  \BibitemOpen
  \bibfield  {author} {\bibinfo {author} {\bibfnamefont {S.~Q.}\ \bibnamefont
  {Lim}}, \bibinfo {author} {\bibfnamefont {A.~J.}\ \bibnamefont {Akey}},
  \bibinfo {author} {\bibfnamefont {E.}~\bibnamefont {Napolitani}}, \bibinfo
  {author} {\bibfnamefont {P.~K.}\ \bibnamefont {Chow}}, \bibinfo {author}
  {\bibfnamefont {J.~M.}\ \bibnamefont {Warrender}},\ and\ \bibinfo {author}
  {\bibfnamefont {J.~S.}\ \bibnamefont {Williams}},\ }\bibfield  {title}
  {\bibinfo {title} {A critical evaluation of {A}g- and {T}i-hyperdoped {S}i
  for {S}i-based infrared light detection},\ }\href
  {https://doi.org/10.1063/5.0035620} {\bibfield  {journal} {\bibinfo
  {journal} {Journal of Applied Physics}\ }\textbf {\bibinfo {volume} {129}},\
  \bibinfo {pages} {065701} (\bibinfo {year} {2021})},\ \Eprint
  {https://arxiv.org/abs/https://doi.org/10.1063/5.0035620}
  {https://doi.org/10.1063/5.0035620} \BibitemShut {NoStop}%
\bibitem [{\citenamefont {Hocine}\ and\ \citenamefont
  {Mathiot}(1988)}]{ho.ma.88}%
  \BibitemOpen
  \bibfield  {author} {\bibinfo {author} {\bibfnamefont {S.}~\bibnamefont
  {Hocine}}\ and\ \bibinfo {author} {\bibfnamefont {D.}~\bibnamefont
  {Mathiot}},\ }\bibfield  {title} {\bibinfo {title} {Titanium diffusion in
  silicon},\ }\href {https://doi.org/10.1063/1.100446} {\bibfield  {journal}
  {\bibinfo  {journal} {Applied Physics Letters}\ }\textbf {\bibinfo {volume}
  {53}},\ \bibinfo {pages} {1269} (\bibinfo {year} {1988})},\ \Eprint
  {https://arxiv.org/abs/https://doi.org/10.1063/1.100446}
  {https://doi.org/10.1063/1.100446} \BibitemShut {NoStop}%
\bibitem [{\citenamefont {Mathiot}\ and\ \citenamefont
  {Barbier}(1991)}]{ma.ba.91}%
  \BibitemOpen
  \bibfield  {author} {\bibinfo {author} {\bibfnamefont {D.}~\bibnamefont
  {Mathiot}}\ and\ \bibinfo {author} {\bibfnamefont {D.}~\bibnamefont
  {Barbier}},\ }\bibfield  {title} {\bibinfo {title} {Solubility enhancement of
  metallic impurities in silicon by rapid thermal annealing},\ }\href
  {https://doi.org/10.1063/1.348444} {\bibfield  {journal} {\bibinfo  {journal}
  {Journal of Applied Physics}\ }\textbf {\bibinfo {volume} {69}},\ \bibinfo
  {pages} {3878} (\bibinfo {year} {1991})},\ \Eprint
  {https://arxiv.org/abs/https://doi.org/10.1063/1.348444}
  {https://doi.org/10.1063/1.348444} \BibitemShut {NoStop}%
\bibitem [{\citenamefont {Olea}\ \emph {et~al.}(2012)\citenamefont {Olea},
  \citenamefont {Pastor}, \citenamefont {García-Hemme}, \citenamefont
  {García-Hernansanz}, \citenamefont {Álvaro {del Prado}}, \citenamefont
  {Mártil},\ and\ \citenamefont {González-Díaz}}]{ol.pa.12}%
  \BibitemOpen
  \bibfield  {author} {\bibinfo {author} {\bibfnamefont {J.}~\bibnamefont
  {Olea}}, \bibinfo {author} {\bibfnamefont {D.}~\bibnamefont {Pastor}},
  \bibinfo {author} {\bibfnamefont {E.}~\bibnamefont {García-Hemme}}, \bibinfo
  {author} {\bibfnamefont {R.}~\bibnamefont {García-Hernansanz}}, \bibinfo
  {author} {\bibnamefont {Álvaro {del Prado}}}, \bibinfo {author}
  {\bibfnamefont {I.}~\bibnamefont {Mártil}},\ and\ \bibinfo {author}
  {\bibfnamefont {G.}~\bibnamefont {González-Díaz}},\ }\bibfield  {title}
  {\bibinfo {title} {Low temperature intermediate band metallic behavior in
  {T}i implanted {S}i},\ }\href
  {https://doi.org/https://doi.org/10.1016/j.tsf.2012.07.014} {\bibfield
  {journal} {\bibinfo  {journal} {Thin Solid Films}\ }\textbf {\bibinfo
  {volume} {520}},\ \bibinfo {pages} {6614} (\bibinfo {year}
  {2012})}\BibitemShut {NoStop}%
\bibitem [{\citenamefont {Akey}\ \emph {et~al.}(2021)\citenamefont {Akey},
  \citenamefont {Mathews},\ and\ \citenamefont {Warrender}}]{ak.ma.21}%
  \BibitemOpen
  \bibfield  {author} {\bibinfo {author} {\bibfnamefont {A.~J.}\ \bibnamefont
  {Akey}}, \bibinfo {author} {\bibfnamefont {J.}~\bibnamefont {Mathews}},\ and\
  \bibinfo {author} {\bibfnamefont {J.~M.}\ \bibnamefont {Warrender}},\
  }\bibfield  {title} {\bibinfo {title} {Maximum {T}i concentrations in {S}i
  quantified with atom probe tomography {(APT)}},\ }\href
  {https://doi.org/10.1063/5.0029981} {\bibfield  {journal} {\bibinfo
  {journal} {Journal of Applied Physics}\ }\textbf {\bibinfo {volume} {129}},\
  \bibinfo {pages} {175701} (\bibinfo {year} {2021})},\ \Eprint
  {https://arxiv.org/abs/https://doi.org/10.1063/5.0029981}
  {https://doi.org/10.1063/5.0029981} \BibitemShut {NoStop}%
\bibitem [{\citenamefont {Dobrosavljevi\ifmmode~\acute{c}\else \'{c}\fi{}}\
  and\ \citenamefont {Kotliar}(1997)}]{do.ko.97}%
  \BibitemOpen
  \bibfield  {author} {\bibinfo {author} {\bibfnamefont {V.}~\bibnamefont
  {Dobrosavljevi\ifmmode~\acute{c}\else \'{c}\fi{}}}\ and\ \bibinfo {author}
  {\bibfnamefont {G.}~\bibnamefont {Kotliar}},\ }\bibfield  {title} {\bibinfo
  {title} {Mean field theory of the {M}ott-{A}nderson transition},\ }\href
  {https://doi.org/10.1103/PhysRevLett.78.3943} {\bibfield  {journal} {\bibinfo
   {journal} {Phys. Rev. Lett.}\ }\textbf {\bibinfo {volume} {78}},\ \bibinfo
  {pages} {3943} (\bibinfo {year} {1997})}\BibitemShut {NoStop}%
\bibitem [{\citenamefont {Dobrosavljevi\'{c}}\ \emph
  {et~al.}(2003)\citenamefont {Dobrosavljevi\'{c}}, \citenamefont {Pastor},\
  and\ \citenamefont {Nikoli\'{c}}}]{do.pa.03}%
  \BibitemOpen
  \bibfield  {author} {\bibinfo {author} {\bibfnamefont {V.}~\bibnamefont
  {Dobrosavljevi\'{c}}}, \bibinfo {author} {\bibfnamefont {A.~A.}\ \bibnamefont
  {Pastor}},\ and\ \bibinfo {author} {\bibfnamefont {B.~K.}\ \bibnamefont
  {Nikoli\'{c}}},\ }\bibfield  {title} {\bibinfo {title} {{Typical medium
  theory of Anderson localization: A local order parameter approach to
  strong-disorder effects}},\ }\href@noop {} {\bibfield  {journal} {\bibinfo
  {journal} {EPL}\ }\textbf {\bibinfo {volume} {62}},\ \bibinfo {pages} {76}
  (\bibinfo {year} {2003})}\BibitemShut {NoStop}%
\bibitem [{\citenamefont {Dobrosavljevi\'{c}}(2010)}]{dobr.10}%
  \BibitemOpen
  \bibfield  {author} {\bibinfo {author} {\bibfnamefont {V.}~\bibnamefont
  {Dobrosavljevi\'{c}}},\ }\bibfield  {title} {\bibinfo {title} {{Typical
  medium theory of Mott-Anderson localization}},\ }\href@noop {} {\bibfield
  {journal} {\bibinfo  {journal} {Int. J. Mod. Phys. B}\ }\textbf {\bibinfo
  {volume} {24}},\ \bibinfo {pages} {1680} (\bibinfo {year}
  {2010})}\BibitemShut {NoStop}%
\bibitem [{\citenamefont {Taylor}(1967)}]{tayl.67}%
  \BibitemOpen
  \bibfield  {author} {\bibinfo {author} {\bibfnamefont {D.}~\bibnamefont
  {Taylor}},\ }\bibfield  {title} {\bibinfo {title} {{Vibrational Properties of
  Imperfect Crystals with Large Defect Concentrations}},\ }\href@noop {}
  {\bibfield  {journal} {\bibinfo  {journal} {Phys. Rev.}\ }\textbf {\bibinfo
  {volume} {156}},\ \bibinfo {pages} {1017} (\bibinfo {year}
  {1967})}\BibitemShut {NoStop}%
\bibitem [{\citenamefont {Soven}(1967)}]{sove.67}%
  \BibitemOpen
  \bibfield  {author} {\bibinfo {author} {\bibfnamefont {P.}~\bibnamefont
  {Soven}},\ }\bibfield  {title} {\bibinfo {title} {{Coherent-Potential Model
  of Substitutional Disordered Alloys}},\ }\href
  {https://doi.org/10.1103/PhysRev.156.809} {\bibfield  {journal} {\bibinfo
  {journal} {Phys. Rev.}\ }\textbf {\bibinfo {volume} {156}},\ \bibinfo {pages}
  {809} (\bibinfo {year} {1967})}\BibitemShut {NoStop}%
\bibitem [{\citenamefont {Jones}(2015)}]{jones.15}%
  \BibitemOpen
  \bibfield  {author} {\bibinfo {author} {\bibfnamefont {R.~O.}\ \bibnamefont
  {Jones}},\ }\bibfield  {title} {\bibinfo {title} {Density functional theory:
  Its origins, rise to prominence, and future},\ }\href
  {https://doi.org/10.1103/RevModPhys.87.897} {\bibfield  {journal} {\bibinfo
  {journal} {Rev. Mod. Phys.}\ }\textbf {\bibinfo {volume} {87}},\ \bibinfo
  {pages} {897} (\bibinfo {year} {2015})}\BibitemShut {NoStop}%
\bibitem [{\citenamefont {Ekuma}\ \emph {et~al.}(2014)\citenamefont {Ekuma},
  \citenamefont {Terletska}, \citenamefont {Tam}, \citenamefont {Meng},
  \citenamefont {Moreno},\ and\ \citenamefont {Jarrell}}]{ek.te.14}%
  \BibitemOpen
  \bibfield  {author} {\bibinfo {author} {\bibfnamefont {C.~E.}\ \bibnamefont
  {Ekuma}}, \bibinfo {author} {\bibfnamefont {H.}~\bibnamefont {Terletska}},
  \bibinfo {author} {\bibfnamefont {K.-M.}\ \bibnamefont {Tam}}, \bibinfo
  {author} {\bibfnamefont {Z.-Y.}\ \bibnamefont {Meng}}, \bibinfo {author}
  {\bibfnamefont {J.}~\bibnamefont {Moreno}},\ and\ \bibinfo {author}
  {\bibfnamefont {M.}~\bibnamefont {Jarrell}},\ }\bibfield  {title} {\bibinfo
  {title} {{Typical medium dynamical cluster approximation for the study of
  Anderson localization in three dimensions}},\ }\href
  {https://doi.org/10.1103/PhysRevB.89.081107} {\bibfield  {journal} {\bibinfo
  {journal} {Phys. Rev. B}\ }\textbf {\bibinfo {volume} {89}},\ \bibinfo
  {pages} {081107} (\bibinfo {year} {2014})}\BibitemShut {NoStop}%
\bibitem [{\citenamefont {Zhang}\ \emph {et~al.}(2015)\citenamefont {Zhang},
  \citenamefont {Terletska}, \citenamefont {Moore}, \citenamefont {Ekuma},
  \citenamefont {Tam}, \citenamefont {Berlijn}, \citenamefont {Ku},
  \citenamefont {Moreno},\ and\ \citenamefont {Jarrell}}]{zh.te.15}%
  \BibitemOpen
  \bibfield  {author} {\bibinfo {author} {\bibfnamefont {Y.}~\bibnamefont
  {Zhang}}, \bibinfo {author} {\bibfnamefont {H.}~\bibnamefont {Terletska}},
  \bibinfo {author} {\bibfnamefont {C.}~\bibnamefont {Moore}}, \bibinfo
  {author} {\bibfnamefont {C.}~\bibnamefont {Ekuma}}, \bibinfo {author}
  {\bibfnamefont {K.-M.}\ \bibnamefont {Tam}}, \bibinfo {author} {\bibfnamefont
  {T.}~\bibnamefont {Berlijn}}, \bibinfo {author} {\bibfnamefont
  {W.}~\bibnamefont {Ku}}, \bibinfo {author} {\bibfnamefont {J.}~\bibnamefont
  {Moreno}},\ and\ \bibinfo {author} {\bibfnamefont {M.}~\bibnamefont
  {Jarrell}},\ }\bibfield  {title} {\bibinfo {title} {Study of multiband
  disordered systems using the typical medium dynamical cluster
  approximation},\ }\href {https://doi.org/10.1103/PhysRevB.92.205111}
  {\bibfield  {journal} {\bibinfo  {journal} {Phys. Rev. B}\ }\textbf {\bibinfo
  {volume} {92}},\ \bibinfo {pages} {205111} (\bibinfo {year}
  {2015})}\BibitemShut {NoStop}%
\bibitem [{\citenamefont {Zhang}\ \emph {et~al.}(2018)\citenamefont {Zhang},
  \citenamefont {Nelson}, \citenamefont {Tam}, \citenamefont {Ku},
  \citenamefont {Yu}, \citenamefont {Vidhyadhiraja}, \citenamefont {Terletska},
  \citenamefont {Moreno}, \citenamefont {Jarrell},\ and\ \citenamefont
  {Berlijn}}]{zh.ne.18}%
  \BibitemOpen
  \bibfield  {author} {\bibinfo {author} {\bibfnamefont {Y.}~\bibnamefont
  {Zhang}}, \bibinfo {author} {\bibfnamefont {R.}~\bibnamefont {Nelson}},
  \bibinfo {author} {\bibfnamefont {K.-M.}\ \bibnamefont {Tam}}, \bibinfo
  {author} {\bibfnamefont {W.}~\bibnamefont {Ku}}, \bibinfo {author}
  {\bibfnamefont {U.}~\bibnamefont {Yu}}, \bibinfo {author} {\bibfnamefont
  {N.~S.}\ \bibnamefont {Vidhyadhiraja}}, \bibinfo {author} {\bibfnamefont
  {H.}~\bibnamefont {Terletska}}, \bibinfo {author} {\bibfnamefont
  {J.}~\bibnamefont {Moreno}}, \bibinfo {author} {\bibfnamefont
  {M.}~\bibnamefont {Jarrell}},\ and\ \bibinfo {author} {\bibfnamefont
  {T.}~\bibnamefont {Berlijn}},\ }\bibfield  {title} {\bibinfo {title} {Origin
  of localization in {T}i-doped {S}i},\ }\href
  {https://doi.org/10.1103/PhysRevB.98.174204} {\bibfield  {journal} {\bibinfo
  {journal} {Phys. Rev. B}\ }\textbf {\bibinfo {volume} {98}},\ \bibinfo
  {pages} {174204} (\bibinfo {year} {2018})}\BibitemShut {NoStop}%
\bibitem [{\citenamefont {Metzner}\ and\ \citenamefont
  {Vollhardt}(1989)}]{me.vo.89}%
  \BibitemOpen
  \bibfield  {author} {\bibinfo {author} {\bibfnamefont {W.}~\bibnamefont
  {Metzner}}\ and\ \bibinfo {author} {\bibfnamefont {D.}~\bibnamefont
  {Vollhardt}},\ }\bibfield  {title} {\bibinfo {title} {{Correlated Lattice
  Fermions in d=$\infty$ dimensions}},\ }\href
  {https://doi.org/10.1103/physrevlett.62.324} {\bibfield  {journal} {\bibinfo
  {journal} {Phys. Rev. Lett.}\ }\textbf {\bibinfo {volume} {62}},\ \bibinfo
  {pages} {324} (\bibinfo {year} {1989})}\BibitemShut {NoStop}%
\bibitem [{\citenamefont {Kotliar}\ and\ \citenamefont
  {Vollhardt}(2004)}]{ko.vo.04}%
  \BibitemOpen
  \bibfield  {author} {\bibinfo {author} {\bibfnamefont {G.}~\bibnamefont
  {Kotliar}}\ and\ \bibinfo {author} {\bibfnamefont {D.}~\bibnamefont
  {Vollhardt}},\ }\bibfield  {title} {\bibinfo {title} {Strongly correlated
  materials: {Insights} from dynamical mean-field theory},\ }\href
  {https://doi.org/10.1063/1.1712502} {\bibfield  {journal} {\bibinfo
  {journal} {Phys. Today}\ }\textbf {\bibinfo {volume} {57}},\ \bibinfo {pages}
  {53} (\bibinfo {year} {2004})}\BibitemShut {NoStop}%
\bibitem [{\citenamefont {\"Ostlin}\ \emph {et~al.}(2020)\citenamefont
  {\"Ostlin}, \citenamefont {Zhang}, \citenamefont {Terletska}, \citenamefont
  {Beiu\ifmmode~\mbox{\c{s}}\else \c{s}\fi{}eanu}, \citenamefont {Popescu},
  \citenamefont {Byczuk}, \citenamefont {Vitos}, \citenamefont {Jarrell},
  \citenamefont {Vollhardt},\ and\ \citenamefont {Chioncel}}]{os.zh.20}%
  \BibitemOpen
  \bibfield  {author} {\bibinfo {author} {\bibfnamefont {A.}~\bibnamefont
  {\"Ostlin}}, \bibinfo {author} {\bibfnamefont {Y.}~\bibnamefont {Zhang}},
  \bibinfo {author} {\bibfnamefont {H.}~\bibnamefont {Terletska}}, \bibinfo
  {author} {\bibfnamefont {F.}~\bibnamefont {Beiu\ifmmode~\mbox{\c{s}}\else
  \c{s}\fi{}eanu}}, \bibinfo {author} {\bibfnamefont {V.}~\bibnamefont
  {Popescu}}, \bibinfo {author} {\bibfnamefont {K.}~\bibnamefont {Byczuk}},
  \bibinfo {author} {\bibfnamefont {L.}~\bibnamefont {Vitos}}, \bibinfo
  {author} {\bibfnamefont {M.}~\bibnamefont {Jarrell}}, \bibinfo {author}
  {\bibfnamefont {D.}~\bibnamefont {Vollhardt}},\ and\ \bibinfo {author}
  {\bibfnamefont {L.}~\bibnamefont {Chioncel}},\ }\bibfield  {title} {\bibinfo
  {title} {Ab initio typical medium theory of substitutional disorder},\ }\href
  {https://doi.org/10.1103/PhysRevB.101.014210} {\bibfield  {journal} {\bibinfo
   {journal} {Phys. Rev. B}\ }\textbf {\bibinfo {volume} {101}},\ \bibinfo
  {pages} {014210} (\bibinfo {year} {2020})}\BibitemShut {NoStop}%
\bibitem [{\citenamefont {\"Ostlin}\ \emph {et~al.}(2018)\citenamefont
  {\"Ostlin}, \citenamefont {Vitos},\ and\ \citenamefont
  {Chioncel}}]{os.vi.18}%
  \BibitemOpen
  \bibfield  {author} {\bibinfo {author} {\bibfnamefont {A.}~\bibnamefont
  {\"Ostlin}}, \bibinfo {author} {\bibfnamefont {L.}~\bibnamefont {Vitos}},\
  and\ \bibinfo {author} {\bibfnamefont {L.}~\bibnamefont {Chioncel}},\
  }\bibfield  {title} {\bibinfo {title} {Correlated electronic structure with
  uncorrelated disorder},\ }\href {https://doi.org/10.1103/PhysRevB.98.235135}
  {\bibfield  {journal} {\bibinfo  {journal} {Phys. Rev. B}\ }\textbf {\bibinfo
  {volume} {98}},\ \bibinfo {pages} {235135} (\bibinfo {year}
  {2018})}\BibitemShut {NoStop}%
\bibitem [{\citenamefont {\"Ostlin}\ \emph {et~al.}(2017)\citenamefont
  {\"Ostlin}, \citenamefont {Vitos},\ and\ \citenamefont
  {Chioncel}}]{os.vi.17}%
  \BibitemOpen
  \bibfield  {author} {\bibinfo {author} {\bibfnamefont {A.}~\bibnamefont
  {\"Ostlin}}, \bibinfo {author} {\bibfnamefont {L.}~\bibnamefont {Vitos}},\
  and\ \bibinfo {author} {\bibfnamefont {L.}~\bibnamefont {Chioncel}},\
  }\bibfield  {title} {\bibinfo {title} {Analytic continuation-free {G}reen's
  function approach to correlated electronic structure calculations},\ }\href
  {https://doi.org/10.1103/PhysRevB.96.125156} {\bibfield  {journal} {\bibinfo
  {journal} {Phys. Rev. B}\ }\textbf {\bibinfo {volume} {96}},\ \bibinfo
  {pages} {125156} (\bibinfo {year} {2017})}\BibitemShut {NoStop}%
\bibitem [{\citenamefont {Bickers}\ and\ \citenamefont
  {Scalapino}(1989)}]{bi.sc.89}%
  \BibitemOpen
  \bibfield  {author} {\bibinfo {author} {\bibfnamefont {N.~E.}\ \bibnamefont
  {Bickers}}\ and\ \bibinfo {author} {\bibfnamefont {D.~J.}\ \bibnamefont
  {Scalapino}},\ }\bibfield  {title} {\bibinfo {title} {Conserving
  approximation for strongly fluctuating electron systems. {I}. formalism and
  calculational approach},\ }\href@noop {} {\bibfield  {journal} {\bibinfo
  {journal} {Ann. Phys. (N. Y.)}\ }\textbf {\bibinfo {volume} {193}},\ \bibinfo
  {pages} {206} (\bibinfo {year} {1989})}\BibitemShut {NoStop}%
\bibitem [{\citenamefont {Lichtenstein}\ and\ \citenamefont
  {Katsnelson}(1998)}]{li.ka.98}%
  \BibitemOpen
  \bibfield  {author} {\bibinfo {author} {\bibfnamefont {A.~I.}\ \bibnamefont
  {Lichtenstein}}\ and\ \bibinfo {author} {\bibfnamefont {M.~I.}\ \bibnamefont
  {Katsnelson}},\ }\bibfield  {title} {\bibinfo {title} {Ab initio calculations
  of quasiparticle band structure in correlated systems: {LDA}++ approach},\
  }\href@noop {} {\bibfield  {journal} {\bibinfo  {journal} {Phys. Rev. B}\
  }\textbf {\bibinfo {volume} {57}},\ \bibinfo {pages} {6884} (\bibinfo {year}
  {1998})}\BibitemShut {NoStop}%
\bibitem [{\citenamefont {Katsnelson}\ and\ \citenamefont
  {Lichtenstein}(1999)}]{ka.li.99}%
  \BibitemOpen
  \bibfield  {author} {\bibinfo {author} {\bibfnamefont {M.~I.}\ \bibnamefont
  {Katsnelson}}\ and\ \bibinfo {author} {\bibfnamefont {A.~I.}\ \bibnamefont
  {Lichtenstein}},\ }\bibfield  {title} {\bibinfo {title} {{LDA}++ approach to
  the electronic structure of magnets: correlation effects in iron},\
  }\href@noop {} {\bibfield  {journal} {\bibinfo  {journal} {J. Phys.: Condens.
  Matter}\ }\textbf {\bibinfo {volume} {11}},\ \bibinfo {pages} {1037}
  (\bibinfo {year} {1999})}\BibitemShut {NoStop}%
\bibitem [{\citenamefont {Pourovskii}\ \emph {et~al.}(2005)\citenamefont
  {Pourovskii}, \citenamefont {Katsnelson},\ and\ \citenamefont
  {Lichtenstein}}]{po.ka.05}%
  \BibitemOpen
  \bibfield  {author} {\bibinfo {author} {\bibfnamefont {L.~V.}\ \bibnamefont
  {Pourovskii}}, \bibinfo {author} {\bibfnamefont {M.~I.}\ \bibnamefont
  {Katsnelson}},\ and\ \bibinfo {author} {\bibfnamefont {A.~I.}\ \bibnamefont
  {Lichtenstein}},\ }\bibfield  {title} {\bibinfo {title} {Correlation effects
  in electronic structure of actinide monochalcogenides},\ }\href@noop {}
  {\bibfield  {journal} {\bibinfo  {journal} {Phys. Rev. B}\ }\textbf {\bibinfo
  {volume} {72}},\ \bibinfo {pages} {115106} (\bibinfo {year}
  {2005})}\BibitemShut {NoStop}%
\bibitem [{\citenamefont {Gr{\aa}n{\"a}s}\ \emph {et~al.}(2012)\citenamefont
  {Gr{\aa}n{\"a}s}, \citenamefont {Di~Marco}, \citenamefont {Thunstr{\"o}m},
  \citenamefont {Nordstr{\"o}m}, \citenamefont {Eriksson}, \citenamefont
  {Bj{\"o}rkman},\ and\ \citenamefont {Wills}}]{gr.ma.12}%
  \BibitemOpen
  \bibfield  {author} {\bibinfo {author} {\bibfnamefont {O.}~\bibnamefont
  {Gr{\aa}n{\"a}s}}, \bibinfo {author} {\bibfnamefont {I.}~\bibnamefont
  {Di~Marco}}, \bibinfo {author} {\bibfnamefont {P.}~\bibnamefont
  {Thunstr{\"o}m}}, \bibinfo {author} {\bibfnamefont {L.}~\bibnamefont
  {Nordstr{\"o}m}}, \bibinfo {author} {\bibfnamefont {O.}~\bibnamefont
  {Eriksson}}, \bibinfo {author} {\bibfnamefont {T.}~\bibnamefont
  {Bj{\"o}rkman}},\ and\ \bibinfo {author} {\bibfnamefont {J.}~\bibnamefont
  {Wills}},\ }\bibfield  {title} {\bibinfo {title} {{Charge self-consistent
  dynamical mean-field theory based on the full-potential linear muffin-tin
  orbital method: Methodology and applications}},\ }\href
  {https://doi.org/10.1016/j.commatsci.2011.11.032} {\bibfield  {journal}
  {\bibinfo  {journal} {Computational Materials Science}\ }\textbf {\bibinfo
  {volume} {55}},\ \bibinfo {pages} {295–302} (\bibinfo {year}
  {2012})}\BibitemShut {NoStop}%
\bibitem [{\citenamefont {Katsnelson}\ \emph {et~al.}(2008)\citenamefont
  {Katsnelson}, \citenamefont {Irkhin}, \citenamefont {Chioncel}, \citenamefont
  {Lichtenstein},\ and\ \citenamefont {de~Groot}}]{ka.ir.08}%
  \BibitemOpen
  \bibfield  {author} {\bibinfo {author} {\bibfnamefont {M.~I.}\ \bibnamefont
  {Katsnelson}}, \bibinfo {author} {\bibfnamefont {V.~Y.}\ \bibnamefont
  {Irkhin}}, \bibinfo {author} {\bibfnamefont {L.}~\bibnamefont {Chioncel}},
  \bibinfo {author} {\bibfnamefont {A.~I.}\ \bibnamefont {Lichtenstein}},\ and\
  \bibinfo {author} {\bibfnamefont {R.~A.}\ \bibnamefont {de~Groot}},\
  }\bibfield  {title} {\bibinfo {title} {Half-metallic ferromagnets: From band
  structure to many-body effects},\ }\href
  {https://doi.org/10.1103/RevModPhys.80.315} {\bibfield  {journal} {\bibinfo
  {journal} {Rev. Mod. Phys.}\ }\textbf {\bibinfo {volume} {80}},\ \bibinfo
  {pages} {315} (\bibinfo {year} {2008})}\BibitemShut {NoStop}%
\bibitem [{\citenamefont {Petukhov}\ \emph {et~al.}(2003)\citenamefont
  {Petukhov}, \citenamefont {Mazin}, \citenamefont {Chioncel},\ and\
  \citenamefont {Lichtenstein}}]{pe.ma.03}%
  \BibitemOpen
  \bibfield  {author} {\bibinfo {author} {\bibfnamefont {A.~G.}\ \bibnamefont
  {Petukhov}}, \bibinfo {author} {\bibfnamefont {I.~I.}\ \bibnamefont {Mazin}},
  \bibinfo {author} {\bibfnamefont {L.}~\bibnamefont {Chioncel}},\ and\
  \bibinfo {author} {\bibfnamefont {A.~I.}\ \bibnamefont {Lichtenstein}},\
  }\bibfield  {title} {\bibinfo {title} {Correlated metals and the {$LDA+U$}
  method},\ }\href@noop {} {\bibfield  {journal} {\bibinfo  {journal} {Phys.
  Rev. B}\ }\textbf {\bibinfo {volume} {67}},\ \bibinfo {pages} {153106}
  (\bibinfo {year} {2003})}\BibitemShut {NoStop}%
\bibitem [{\citenamefont {Vidberg}\ and\ \citenamefont
  {Serene}(1977)}]{vi.se.77}%
  \BibitemOpen
  \bibfield  {author} {\bibinfo {author} {\bibfnamefont {H.~J.}\ \bibnamefont
  {Vidberg}}\ and\ \bibinfo {author} {\bibfnamefont {J.~W.}\ \bibnamefont
  {Serene}},\ }\bibfield  {title} {\bibinfo {title} {Solving the {E}liashberg
  equations by means of {N}-point {P}ad\'e approximants},\ }\href@noop {}
  {\bibfield  {journal} {\bibinfo  {journal} {J. Low Temp. Phys.}\ }\textbf
  {\bibinfo {volume} {29}},\ \bibinfo {pages} {179} (\bibinfo {year}
  {1977})}\BibitemShut {NoStop}%
\bibitem [{\citenamefont {\"Ostlin}\ \emph {et~al.}(2012)\citenamefont
  {\"Ostlin}, \citenamefont {Chioncel},\ and\ \citenamefont
  {Vitos}}]{os.ch.12}%
  \BibitemOpen
  \bibfield  {author} {\bibinfo {author} {\bibfnamefont {A.}~\bibnamefont
  {\"Ostlin}}, \bibinfo {author} {\bibfnamefont {L.}~\bibnamefont {Chioncel}},\
  and\ \bibinfo {author} {\bibfnamefont {L.}~\bibnamefont {Vitos}},\ }\bibfield
   {title} {\bibinfo {title} {One-particle spectral function and analytic
  continuation for many-body implementation in the exact muffin-tin orbitals
  method},\ }\href {https://doi.org/10.1103/PhysRevB.86.235107} {\bibfield
  {journal} {\bibinfo  {journal} {Phys. Rev. B}\ }\textbf {\bibinfo {volume}
  {86}},\ \bibinfo {pages} {235107} (\bibinfo {year} {2012})}\BibitemShut
  {NoStop}%
\bibitem [{\citenamefont {Marinopoulos}\ \emph {et~al.}(2015)\citenamefont
  {Marinopoulos}, \citenamefont {Santos},\ and\ \citenamefont
  {Coutinho}}]{ma.sa.15}%
  \BibitemOpen
  \bibfield  {author} {\bibinfo {author} {\bibfnamefont {A.~G.}\ \bibnamefont
  {Marinopoulos}}, \bibinfo {author} {\bibfnamefont {P.}~\bibnamefont
  {Santos}},\ and\ \bibinfo {author} {\bibfnamefont {J.}~\bibnamefont
  {Coutinho}},\ }\bibfield  {title} {\bibinfo {title} {{DFT}+${U}$ study of
  electrical levels and migration barriers of early $3d$ and $4d$ transition
  metals in silicon},\ }\href {https://doi.org/10.1103/PhysRevB.92.075124}
  {\bibfield  {journal} {\bibinfo  {journal} {Phys. Rev. B}\ }\textbf {\bibinfo
  {volume} {92}},\ \bibinfo {pages} {075124} (\bibinfo {year}
  {2015})}\BibitemShut {NoStop}%
\bibitem [{\citenamefont {Perdew}\ \emph {et~al.}(1996)\citenamefont {Perdew},
  \citenamefont {Burke},\ and\ \citenamefont {Ernzerhof}}]{pe.bu.96}%
  \BibitemOpen
  \bibfield  {author} {\bibinfo {author} {\bibfnamefont {J.~P.}\ \bibnamefont
  {Perdew}}, \bibinfo {author} {\bibfnamefont {K.}~\bibnamefont {Burke}},\ and\
  \bibinfo {author} {\bibfnamefont {M.}~\bibnamefont {Ernzerhof}},\ }\bibfield
  {title} {\bibinfo {title} {{Generalized Gradient Approximation Made
  Simple}},\ }\href {https://doi.org/10.1103/PhysRevLett.77.3865} {\bibfield
  {journal} {\bibinfo  {journal} {Phys. Rev. Lett.}\ }\textbf {\bibinfo
  {volume} {77}},\ \bibinfo {pages} {3865} (\bibinfo {year}
  {1996})}\BibitemShut {NoStop}%
\bibitem [{\citenamefont {Bullen}(2003)}]{bullen.03}%
  \BibitemOpen
  \bibfield  {author} {\bibinfo {author} {\bibfnamefont {P.~S.}\ \bibnamefont
  {Bullen}},\ }\href@noop {} {\emph {\bibinfo {title} {Handbook of Means and
  Their Inequalities}}}\ (\bibinfo  {publisher} {Kluwer Academic Publishers},\
  \bibinfo {address} {Dordrecht},\ \bibinfo {year} {2003})\BibitemShut
  {NoStop}%
\bibitem [{\citenamefont {Schubert}\ \emph {et~al.}(2010)\citenamefont
  {Schubert}, \citenamefont {Schleede}, \citenamefont {Byczuk}, \citenamefont
  {Fehske},\ and\ \citenamefont {Vollhardt}}]{sc.sc.10}%
  \BibitemOpen
  \bibfield  {author} {\bibinfo {author} {\bibfnamefont {G.}~\bibnamefont
  {Schubert}}, \bibinfo {author} {\bibfnamefont {J.}~\bibnamefont {Schleede}},
  \bibinfo {author} {\bibfnamefont {K.}~\bibnamefont {Byczuk}}, \bibinfo
  {author} {\bibfnamefont {H.}~\bibnamefont {Fehske}},\ and\ \bibinfo {author}
  {\bibfnamefont {D.}~\bibnamefont {Vollhardt}},\ }\bibfield  {title} {\bibinfo
  {title} {Distribution of the local density of states as a criterion for
  {A}nderson localization: Numerically exact results for various lattices in
  two and three dimensions},\ }\href
  {https://doi.org/10.1103/PhysRevB.81.155106} {\bibfield  {journal} {\bibinfo
  {journal} {Phys. Rev. B}\ }\textbf {\bibinfo {volume} {81}},\ \bibinfo
  {pages} {155106} (\bibinfo {year} {2010})}\BibitemShut {NoStop}%
\bibitem [{\citenamefont {Markevich}\ \emph {et~al.}(2014)\citenamefont
  {Markevich}, \citenamefont {Leonard}, \citenamefont {Peaker}, \citenamefont
  {Hamilton}, \citenamefont {Marinopoulos},\ and\ \citenamefont
  {Coutinho}}]{ma.le.14}%
  \BibitemOpen
  \bibfield  {author} {\bibinfo {author} {\bibfnamefont {V.~P.}\ \bibnamefont
  {Markevich}}, \bibinfo {author} {\bibfnamefont {S.}~\bibnamefont {Leonard}},
  \bibinfo {author} {\bibfnamefont {A.~R.}\ \bibnamefont {Peaker}}, \bibinfo
  {author} {\bibfnamefont {B.}~\bibnamefont {Hamilton}}, \bibinfo {author}
  {\bibfnamefont {A.~G.}\ \bibnamefont {Marinopoulos}},\ and\ \bibinfo {author}
  {\bibfnamefont {J.}~\bibnamefont {Coutinho}},\ }\bibfield  {title} {\bibinfo
  {title} {Titanium in silicon: Lattice positions and electronic properties},\
  }\href {https://doi.org/10.1063/1.4871702} {\bibfield  {journal} {\bibinfo
  {journal} {Appl. Phys. Lett.}\ }\textbf {\bibinfo {volume} {104}},\ \bibinfo
  {pages} {152105} (\bibinfo {year} {2014})},\ \Eprint
  {https://arxiv.org/abs/https://doi.org/10.1063/1.4871702}
  {https://doi.org/10.1063/1.4871702} \BibitemShut {NoStop}%
\bibitem [{\citenamefont {Berthod}\ \emph {et~al.}(2013)\citenamefont
  {Berthod}, \citenamefont {Mravlje}, \citenamefont {Deng}, \citenamefont
  {\ifmmode~\check{Z}\else \v{Z}\fi{}itko}, \citenamefont {van~der Marel},\
  and\ \citenamefont {Georges}}]{be.mr.13}%
  \BibitemOpen
  \bibfield  {author} {\bibinfo {author} {\bibfnamefont {C.}~\bibnamefont
  {Berthod}}, \bibinfo {author} {\bibfnamefont {J.}~\bibnamefont {Mravlje}},
  \bibinfo {author} {\bibfnamefont {X.}~\bibnamefont {Deng}}, \bibinfo {author}
  {\bibfnamefont {R.}~\bibnamefont {\ifmmode~\check{Z}\else \v{Z}\fi{}itko}},
  \bibinfo {author} {\bibfnamefont {D.}~\bibnamefont {van~der Marel}},\ and\
  \bibinfo {author} {\bibfnamefont {A.}~\bibnamefont {Georges}},\ }\bibfield
  {title} {\bibinfo {title} {Non-{D}rude universal scaling laws for the optical
  response of local {F}ermi liquids},\ }\href
  {https://doi.org/10.1103/PhysRevB.87.115109} {\bibfield  {journal} {\bibinfo
  {journal} {Phys. Rev. B}\ }\textbf {\bibinfo {volume} {87}},\ \bibinfo
  {pages} {115109} (\bibinfo {year} {2013})}\BibitemShut {NoStop}%
\bibitem [{\citenamefont {Chubukov}\ and\ \citenamefont
  {Maslov}(2012)}]{ch.ma.12}%
  \BibitemOpen
  \bibfield  {author} {\bibinfo {author} {\bibfnamefont {A.~V.}\ \bibnamefont
  {Chubukov}}\ and\ \bibinfo {author} {\bibfnamefont {D.~L.}\ \bibnamefont
  {Maslov}},\ }\bibfield  {title} {\bibinfo {title}
  {First-{M}atsubara-frequency rule in a {F}ermi liquid. {I}. {F}ermionic
  self-energy},\ }\href {https://doi.org/10.1103/PhysRevB.86.155136} {\bibfield
   {journal} {\bibinfo  {journal} {Phys. Rev. B}\ }\textbf {\bibinfo {volume}
  {86}},\ \bibinfo {pages} {155136} (\bibinfo {year} {2012})}\BibitemShut
  {NoStop}%
\end{thebibliography}%
\end{document}